\documentclass[aps,twocolumn,showpacs]{revtex4}
\usepackage{graphics,epsfig}

\usepackage{amsmath}
\usepackage{amssymb}
\begin{document}
\title{Exact Relativistic Static Charged Perfect Fluid Disks}
\author{D. Vogt\footnote{e-mail: danielvt@ifi.unicamp.br}}

\affiliation{
Instituto de F\'{\i}sica Gleb Wataghin, Universidade Estadual de Campinas
13083-970 Campinas, S.\ P., Brazil}

\author{
P. S. Letelier\footnote{e-mail: letelier@ime.unicamp.br}}
\affiliation{
Departamento de Matem\'{a}tica Aplicada-IMECC, Universidade Estadual
de Campinas 13083-970 Campinas, S.\ P., Brazil}

\begin{abstract}
Using the well-known ``displace, cut and reflect'' method used to generate 
disks from given solutions of Einstein field equations, we construct static 
charged disks made of perfect fluid based on the Reissner-Nordstr\"{o}m
solution in isotropic coordinates. We also derive a simple stability condition for charged and 
non charged perfect fluid disks. As expected, we find that the presence of charge increases 
the regions of instability of the disks.
\end{abstract}
\maketitle
\setlength{\parindent}{3em}

\section{Introduction}
Axisymmetric solutions of Einstein field equations corresponding to disklike configurations of 
matter have been extensively studied. These solutions can be static or stationary and with or 
without radial pressure. Solutions for static thin disks without radial pressure were first 
studied by Bonnor and Sackfield \cite{Bonnor1}, and Morgan and Morgan \cite{Morgan1}, and with 
radial pressure by Morgan and Morgan \cite{Morgan2}. Other classes of static thin disk solutions
have been obtained \cite{Lynden1}--\cite{Bicak2}, while stationary thin disks were studied in \cite{Bicak3}--\cite{Gonzalez1}.
Also thin disks with radial tension \cite{Gonzalez2}, magnetic fields \cite{Letelier1} and both electric 
and magnetic fields \cite{Katz1} have been studied.  An exact solution to the problem of a rigidly 
rotating disk of dust in terms of ultraelliptic functions was reported in \cite{Neugebauer}, while models of 
static relativistic counterrotating dust disks were recently presented in \cite{Gonzalez3}. Structures of 
nonaxisymmetric matter distributions of static charged dust were also recently studied \cite{Vogt2}. The non-linear
superposition of a disk and a black hole was first considered by Lemos and Letelier \cite{Lemos2}.
Models of thin disks and thin disks with halos made of perfect fluids were considered in \cite{Vogt}.
The generalization of the ``displace, cut and reflect'' method (Sec.\ \ref{sec_form}) to construct thick
static disks was considered by Gonz\'{a}lez and Letelier \cite{Gonzalez4}.
For a recent review on relativistic accretion disks see \cite{Karas}.

Gravitationally bound stellar objects are likely to be positively charged due to the fact that the electrons are lighter than the protons so the formers 
can more easily scape from the stellar object. One should expect to have an
 equilibrium situation  when the electrostatic energy of the electron ($eV$) is of the order of its thermal energy ($kT$) \cite{Bally}, then  
the scape would stop. Similar considerations can be found in Ref.\ \cite{Shvartsman}.
 
The aim of this paper is to present a charged version of perfect fluid
disks that were studied in \cite{Vogt}. They are constructed with the well known ``displace,
cut and reflect'' method applied to the Reissner-Nordstr\"{o}m metric in isotropic coordinates.  The stability of these new model of charged disks will also be considered. We present and extension of the
Rayleigh criteria of stability \cite{Rayleigh} for the case of relativistic disks made of a
 charged perfect fluid. We find a simple condition to have gravitationally bounded disks.

 The paper is divided as
follows. Sec.\ \ref{sec_form} discusses the formalism which can be
used to construct disks given a solution of Einstein-Maxwell equations, and the main 
physical variables of the disk. In Sec.\ \ref{sec_stab}   Rayleigh inspired 
criteria of stability criteria for
the stability of charged and noncharged perfect fluid disks is established. 
The properties and stability of the charged disks
are discussed in Sec.\ \ref{sec_disk}. Finally, in Sec.\ \ref{sec_disc}, we 
discuss our results and make some considerations about the construction 
of charged perfect fluid disks with halos. The ``displace, cut and reflect'' method
is reviewed in an Appendix.

\section{Einstein-Maxwell Equations and Disks} \label{sec_form}

The isotropic metric representing a static spherically symmetric space-time
can be expressed as
\begin{equation} \label{eq_metric1}
\mathrm{d}s^2=e^{\nu(r)}\mathrm{d}t^2-e^{\lambda(r)}(
\mathrm{d}r^2+r^2 d\Omega^2),\end{equation}
where $d\Omega^2=\mathrm{d}\theta^2+\sin^2 \theta \mathrm{d}\varphi^2$. The same 
metric in cylindrical coordinates $(t,R,z,\varphi)$ reads:
\begin{equation} \label{eq_metric2}
\mathrm{d}s^2=e^{\nu(R,z)}\mathrm{d}t^2-e^{\lambda(R,z)}\left(
\mathrm{d}R^2+\mathrm{d}z^2+R^2 \mathrm{d}\varphi^2 \right) \mbox{.}
\end{equation}

The Einstein-Maxwell system of equations is given by
\begin{align}
G_{\mu \nu} &=8\pi T_{\mu \nu} \mbox{,} \\
T_{\mu \nu} &=\frac{1}{4\pi}\left( F_{\mu}{}^{\sigma}F_{\nu \sigma}+
\frac{1}{4}g_{\mu\nu}F_{\rho\sigma}F^{\rho\sigma}\right) \mbox{,} \label{eq_elm_ten}\\
F^{\mu\nu}{}_{;\mu} &=0 \mbox{,}\\
F_{\mu\nu}&=A_{\nu,\mu}-A_{\mu,\nu} \mbox{,}
\end{align}
where all symbols have their usual meaning. We use geometric units with $G=c=1$.

The method used to generate the metric of the disk and its 
material content is the well known ``displace, cut and reflect'' method (see the Appendix)
that was first used by Kuzmin \cite{Kuzmin} and Toomre \cite{Toomre}
to construct Newtonian models of disks, and later extended to general
relativity (see, for example \cite{Bicak1,Gonzalez1}).
Given a solution of the Einstein-Maxwell equation, this procedure is 
mathematically equivalent to apply the transformation
$z \rightarrow |z|+a$, with $a$ constant, on that solution. In the Einstein tensor we
have first and second derivatives of $z$.  Remembering that $\partial_z |z|=2
\vartheta(z)-1$ and $\partial_{zz} |z|=2\delta(z)$, where
$\vartheta(z)$ and $\delta(z)$ are, respectively, the Heaviside
function and the Dirac distribution, Einstein-Maxwell equations give us
\begin{align}
G_{\mu \nu} &=8\pi (T_{\mu \nu}^{\mathrm{elm.}}+Q_{\mu\nu}\delta(z)) \mbox{,} 
\label{eq_eins_disk} \\
F^{\mu\nu}{}_{;\mu} &=4 \pi J^{\nu}\delta(z) \mbox{,} \label{eq_max_disk}
\end{align}
where $T_{\mu \nu}^{\mathrm{elm.}}$ is the electromagnetic tensor Eq.\ (\ref{eq_elm_ten}),
$Q_{\mu\nu}$ is the energy-momentum tensor on the plane $z=0$ and $J^{\nu}$ is 
the current density on the plane $z=0$. For the metric (\ref{eq_metric2}), the non-zero 
components of $Q_{\mu\nu}$ are
\begin{align}
Q^t_t &=\frac{1}{16 \pi} \left[
-b^{zz}+g^{zz}(b^R_R+b^z_z+b^{\varphi}_{\varphi}) \right] \mbox{,}
\label{eq_qtt}\\ 
Q^R_R &= Q^{\varphi}_{\varphi} =\frac{1}{16 \pi}
\left[ -b^{zz}+g^{zz}(b^t_t+b^R_R+b^z_z) \right] \mbox{,}
\label{eq_qrr}
\end{align}
where $b_{\mu\nu}$ denote the jump of the first derivatives of the metric
tensor on the plane $z=0$,
\begin{equation}
b_{\mu\nu}=g_{\mu\nu,z}|_{z=0^+}-g_{\mu\nu,z}|_{z=0^-} \mbox{,}
\end{equation}
and the other quantities are evaluated at $z=0^+$. The electromagnetic potential
for an electric field is
\begin{equation} \label{eq_elec_pot}
A_{\mu}=(\phi,0,0,0) \mbox{.}
\end{equation}
Using Eq.\ (\ref{eq_elec_pot}) and Eq.\ (\ref{eq_max_disk}), the only non-zero
component of the current density on the plane $z=0$ is
\begin{equation} \label{eq_curr_disk}
J^t=\frac{1}{4\pi}g^{zz}g^{tt}a_t \mbox{,}
\end{equation}
where $a_{\mu}$ denote the jump of the first derivatives of the electromagnetic 
potential on the plane $z=0$,
\begin{equation}
a_{\mu}=A_{\mu,z}|_{z=0^+}-A_{\mu,z}|_{z=0^-} \mbox{,}
\end{equation}
and the other quantities are evaluated at $z=0^+$. The ``physical measure'' of
length in the direction $\partial_z$ for metric (\ref{eq_metric2}) is $\sqrt{-g_{zz}}$, 
then the invariant distribution is $\delta(z)/\sqrt{-g_{zz}}$. Thus the ``true'' surface
energy density $\sigma$ and the azimuthal and radial pressures or tensions $(P)$
are:
\begin{equation} \label{eq_disk_surf}
\sigma=\sqrt{-g_{zz}}Q^t_t \text{,} \quad
P=-\sqrt{-g_{zz}}Q^R_R=-\sqrt{-g_{zz}}Q^\varphi_\varphi \mbox{.}
\end{equation}
Since $J^{\mu}=\rho U^{\mu}$, where $U^{\mu}=\delta^{\mu}_t/\sqrt{g_{tt}}$,
the ``true'' surface charge density $\rho$ is 
\begin{equation} \label{eq_disk_ch}
\rho=\sqrt{-g_{zz}g_{tt}} J^t \mbox{.}
\end{equation}
\section{Stability Conditions for Perfect Fluid Disks} \label{sec_stab}

One way to explain stability of static disks without radial pressure is to assume 
that the particles on the disk plane move under the action of their own gravitational 
field in such a way that as many particles move clockwise as counterclockwise
(counterrotating model). With this assumption, stability of the matter on the disk 
can be associated with stability of circular orbits along geodesics (see \cite{Letelier2} for 
a detailed explanation). The usual stability criteria for circular orbits is adapted from 
the Rayleigh criteria of stability for a rotating fluid \cite{Rayleigh}. One finds that stability against
small radial perturbations is achieved when
\begin{equation}
h\frac{dh}{dr}>0 \mbox{,}
\end{equation}
where $h$ is the specific angular momentum of the circular orbit.

In the case of perfect fluid disks the situation is somewhat different, since radial
pressure can equilibrate the inward gravitational force and no counterrotating hypothesis 
is needed. Using $T^{\mu \nu}{}_{;\nu}=0$
where $T^{\mu \nu}$ is the sum of the energy-momentum tensor for a perfect fluid and Eq.\ (\ref{eq_elm_ten}), 
the equilibrium condition in the radial direction on the plane $z=0$ is given by
\begin{equation} \label{eq_equil3}
\frac{1}{2}(P+\sigma)(e^{\nu})_{,R}=-\rho e^{\nu/2}\phi_{,R}-e^{\nu}P_{,R} \mbox{.}
\end{equation}
The left side of Eq.\ (\ref{eq_equil3}) can be interpreted as the gravitational force which 
equilibrates the pressure and electric forces that appear on the right side. Now suppose
an element of fluid at radius $R$ is displaced to $R+\Delta R$ keeeping $P$, $\sigma$ and $\rho$
constant. The right side of Eq.\ (\ref{eq_equil3}) becomes
\begin{equation} \label{eq_equil4}
-e^{\nu(R+\Delta R)/2}\phi_{,R}(R+\Delta R)\rho(R)-e^{\nu(R+\Delta R)}P_{,R}(R) \mbox{.}
\end{equation}
These ``forces'' should be compared with the right side of Eq.\ (\ref{eq_equil3}) at radius $R+\Delta R$:
\begin{eqnarray} \label{eq_equil5}
-e^{\nu(R+\Delta R)/2}\phi_{,R}(R+\Delta R)\rho(R+\Delta R) \nonumber \\-e^{\nu(R+\Delta R)}P_{,R}(R+\Delta R) \mbox{.}
\end{eqnarray}
To have stability expression (\ref{eq_equil4}) must be less then expression (\ref{eq_equil5}). Expanding
$\rho(R+\Delta R)$ and $P_{,R}(R+\Delta R)$ around $R$, we get
\begin{equation} \label{eq_est_fluid}
e^{\nu/2}\phi_{,R}\rho_{,R}+e^{\nu}P_{,RR}<0 \mbox{.}
\end{equation}
For an uncharged fluid, condition (\ref{eq_est_fluid}) reduces to $P_{,RR}<0$.
 
Note that this criterium of stability refers only to radial perturbations
 of the pressure and charge. It really gives us a condition to have gravitationally bounded systems. In 
other  words the disks do not explode, but they can collapse. In the general case, the study stability of 
gravitating  systems reduces to the much harder problem of 
 the study of the eigenvalue problem for a nontrivial elliptic operator 
\cite{Friedman}. 
\section{Charged Perfect Fluid Disks} \label{sec_disk}

We apply now the results of the previous sections to construct charged disks.The Reissner-Nordstr\"{o}m 
solution in Schwarzschild coordinates is given by

\
\begin{equation}
\mathrm{d}s^2=\left( 1-\frac{2m}{r}+\frac{Q^2}{r^2} \right) \mathrm{d}t^2-
\frac{\mathrm{d}r^2}{\left( 1-\frac{2m}{r}+\frac{Q^2}{r^2} \right)}-r^2d\Omega^2 \mbox{,} 
\label{eq_reiss1}
\end{equation}
where $m$ and $Q$ are, respectively, the mass and charge of the black hole, 
and $m>Q$. The electromagnetic potential associated to solution (\ref{eq_reiss1}) is
\begin{equation} \label{eq_pot_c1}
A_{\mu}=\left(\frac{Q}{r},0,0,0 \right) \mbox{.}
\end{equation}
With the radial coordinate transformation 
\begin{equation}
r=r' \left(1+\frac{m+Q}{2r'}\right)\left(1+\frac{m-Q}{2r'}\right) \mbox{,}
\end{equation}
metric (\ref{eq_reiss1}) and Eq.\ (\ref{eq_pot_c1}) can be expressed in isotropic coordinates $(t,r',\theta,\varphi)$ as
\begin{widetext}
\begin{equation}
\mathrm{d}s^2 =\frac{\left[ 1-\frac{(m^2-Q^2)}{4r'^2} \right]^2}{\left[ 1+\frac{(m+Q)}{2r'} \right]^2
\left[ 1+\frac{(m-Q)}{2r'} \right]^2} \mathrm{d}t^2 -
\left[ 1+\frac{(m+Q)}{2r'} \right]^2 \left[ 1+\frac{(m-Q)}{2r'} \right]^2  
 (\mathrm{d}r'^2+r'^2 \mathrm{d}\theta^2+
r'^2\sin^2 \theta \mathrm{d} \varphi^2) \mbox{,} \label{eq_reiss2} 
\end{equation}
\end{widetext}
\begin{equation}
A_{\mu}= \left( \frac{Q}{r'\left(1+\frac{m+Q}{2r'}\right)\left(1+\frac{m-Q}{2r'}\right)},0,0,0 \right)
 \label{eq_pot_c2} \mbox{.}
\end{equation}
Transforming Eq.\ (\ref{eq_reiss2})-(\ref{eq_pot_c2}) to cylindrical coordinates, and using
Eq.\ (\ref{eq_disk_surf}) and (\ref{eq_disk_ch}), we obtain
a disk with surface energy density $\sigma=\bar{\sigma}/m$, equal radial and azimuthal pressures
(or tensions) $P=\bar{P}/m$ and surface charge density $\rho=\bar{\rho}/m$ where
\begin{align}
\bar{\sigma} &=\frac{4\tilde{a}}{\pi}\frac{2\sqrt{\tilde{R}^2+\tilde{a}^2}+1-\tilde{Q}^2}
{[(1+2\sqrt{\tilde{R}^2+\tilde{a}^2})^2-\tilde{Q}^2]^2} \mbox{,} \label{eq_sigma_r} \\
\bar{P} &=-\frac{2\tilde{a}}{\pi}\frac{1-\tilde{Q}^2}{[(1+2\sqrt{\tilde{R}^2+\tilde{a}^2})^2-\tilde{Q}^2]
[1-\tilde{Q}^2-4(\tilde{R}^2+\tilde{a}^2)]} \mbox{,}\label{eq_P_r} \\
\bar{\rho} &=\frac{8\tilde{Q}\tilde{a}\sqrt{\tilde{R}^2+\tilde{a}^2}}
{\pi\left[ 4\sqrt{\tilde{R}^2+\tilde{a}^2}(1+\sqrt{\tilde{R}^2+\tilde{a}^2})+1-\tilde{Q}^2 \right]^2} \mbox{,} \label{eq_rho_r}
\end{align}
with $\tilde{R}=R/m$, $\tilde{a}=a/m$ and $\tilde{Q}=Q/m$.

Eq.\ (\ref{eq_sigma_r}) shows that the disk's surface density is always positive (weak energy condition)
for $\tilde{Q}<1$. Positive values (pression) for the stresses in azimuthal and radial directions
are obtained if $\tilde{a}>\sqrt{1-\tilde{Q}^2}/2$. The velocity of sound propagation $V$, defined 
as $V^2=\frac{dP}{d\sigma}$, is calculated using Eq.\ (\ref{eq_sigma_r}) and Eq.\ (\ref{eq_P_r}):
\begin{widetext}
\begin{equation} \label{eq_V_r}
V^2 =\frac{(1-\tilde{Q}^2)[(1+2\sqrt{\tilde{R}^2+\tilde{a}^2})^2-\tilde{Q}^2][(1+2\sqrt{\tilde{R}^2+\tilde{a}^2})^2
(1-4\sqrt{\tilde{R}^2+\tilde{a}^2})-\tilde{Q}^2]}{[1-\tilde{Q}^2-4(\tilde{R}^2+\tilde{a}^2)]^2
[-3(1+2\sqrt{\tilde{R}^2+\tilde{a}^2})^2+\tilde{Q}^2(3+8\sqrt{\tilde{R}^2+\tilde{a}^2})]} \mbox{.}
\end{equation}
\end{widetext}

\begin{figure}
\centering
\includegraphics[scale=0.6]{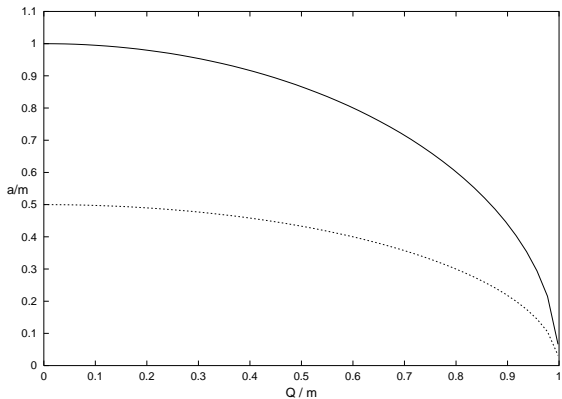}
\caption{Curve $V^2=1$ (solid curve)  as a function  of the parameters
$\tilde{a}=a/m$ and $\tilde{Q}=Q/m$ for the charged disk of perfect fluid. 
 Also $\tilde{a}=\sqrt{1-\tilde{Q}^2}/2$ (dotted curve).} \label{fig1}
\end{figure}

Fig.\ \ref{fig1} shows the curves of $V^2=1$ (solid curve) and of $\tilde{a}=\sqrt{1-\tilde{Q}^2}/2$ (dotted curve) where $\bar{P}$
changes sign as functions of
the parameters $\tilde{a}=a/m$ and $\tilde{Q}=Q/m$. Above the dotted curve, stresses are
positive (pressure) for all $\tilde{R}$ and above the solid curve, condition $ V^2<1$ is also
satisfied for all $\tilde{R}$. Thus, choosing values for $\tilde{a}$ and $\tilde{Q}$ that lie above the solid curve
ensures that the entire disk will have pressures and sublumial sound velocities. We also
note that with increasing charge the range of the cut parameter $a$ that generates disks for which the conditions stated above
are satisfied is enlarged.

\begin{figure}
\centering
\includegraphics[scale=0.6]{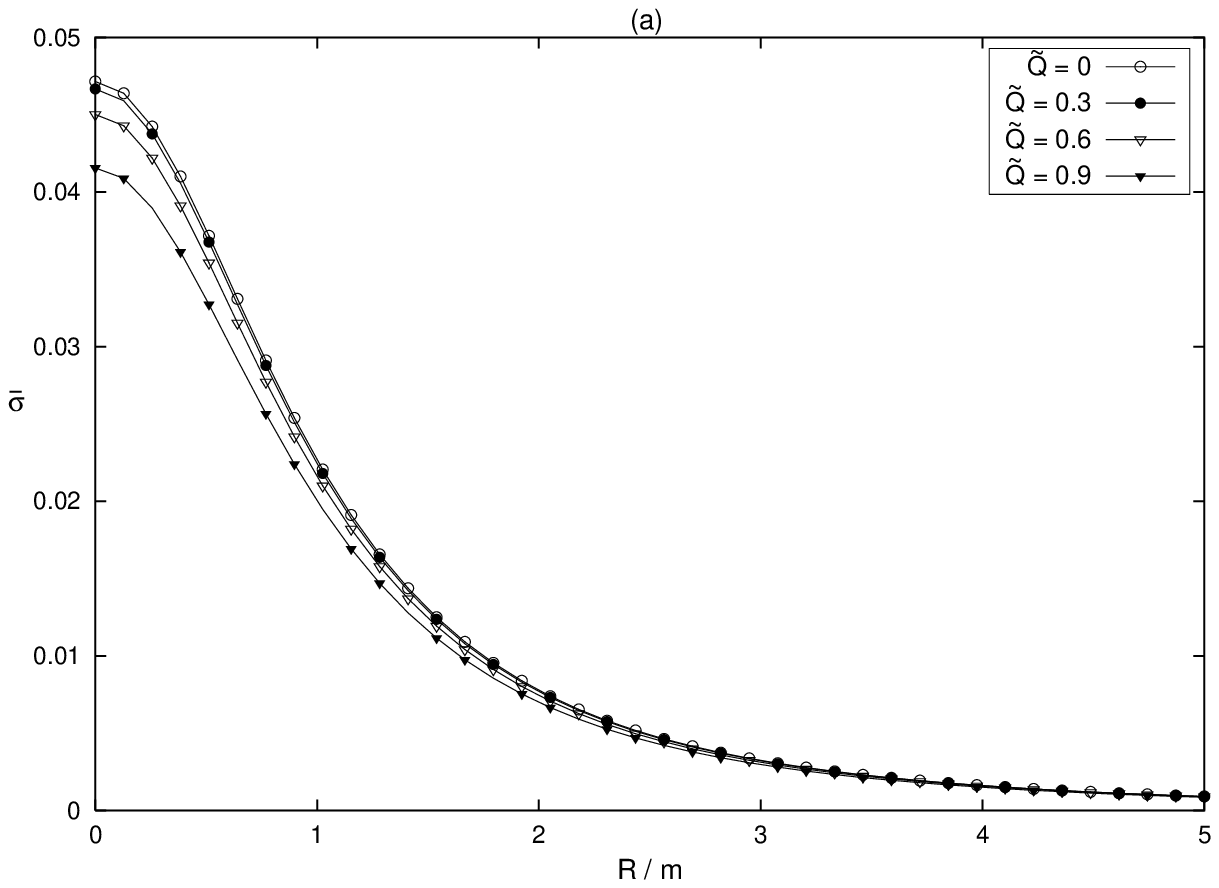}
\includegraphics[scale=0.6]{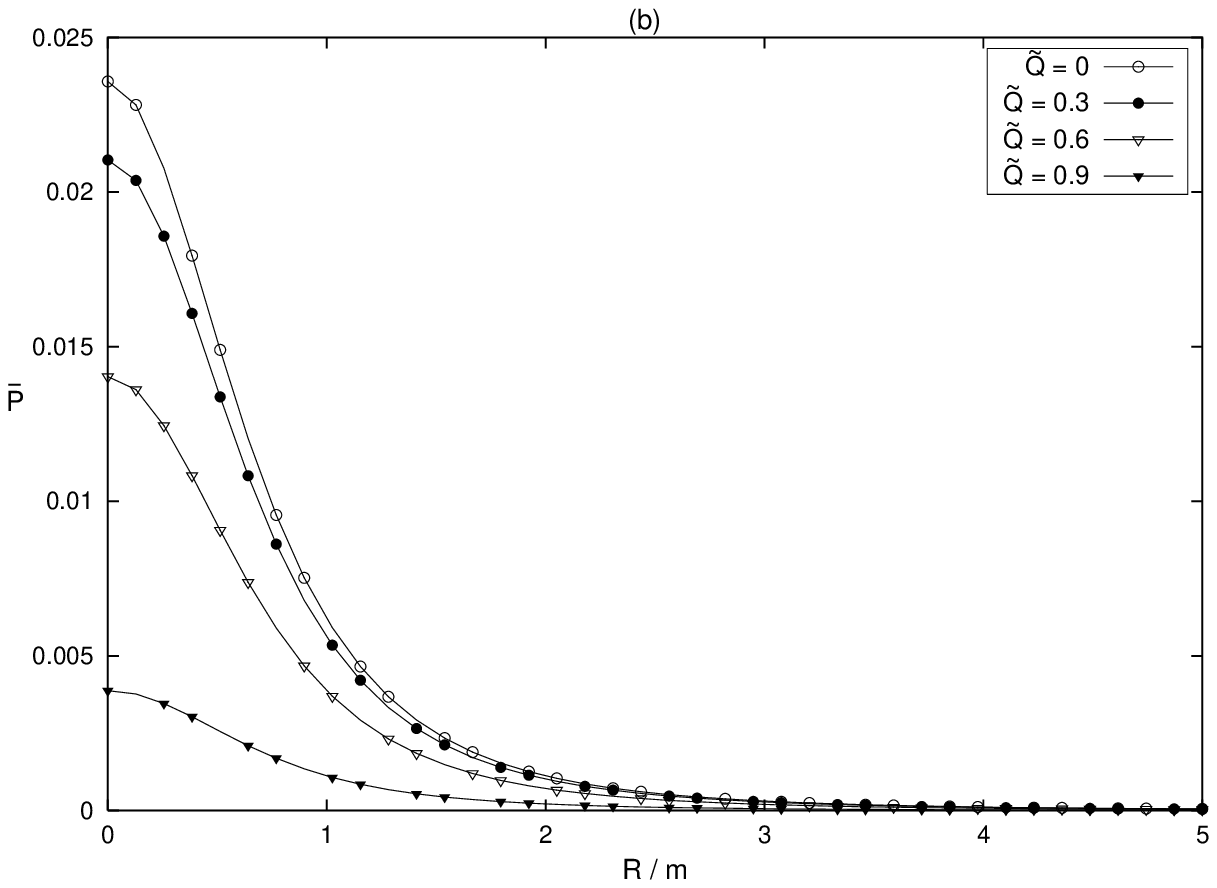}

\includegraphics[scale=0.6]{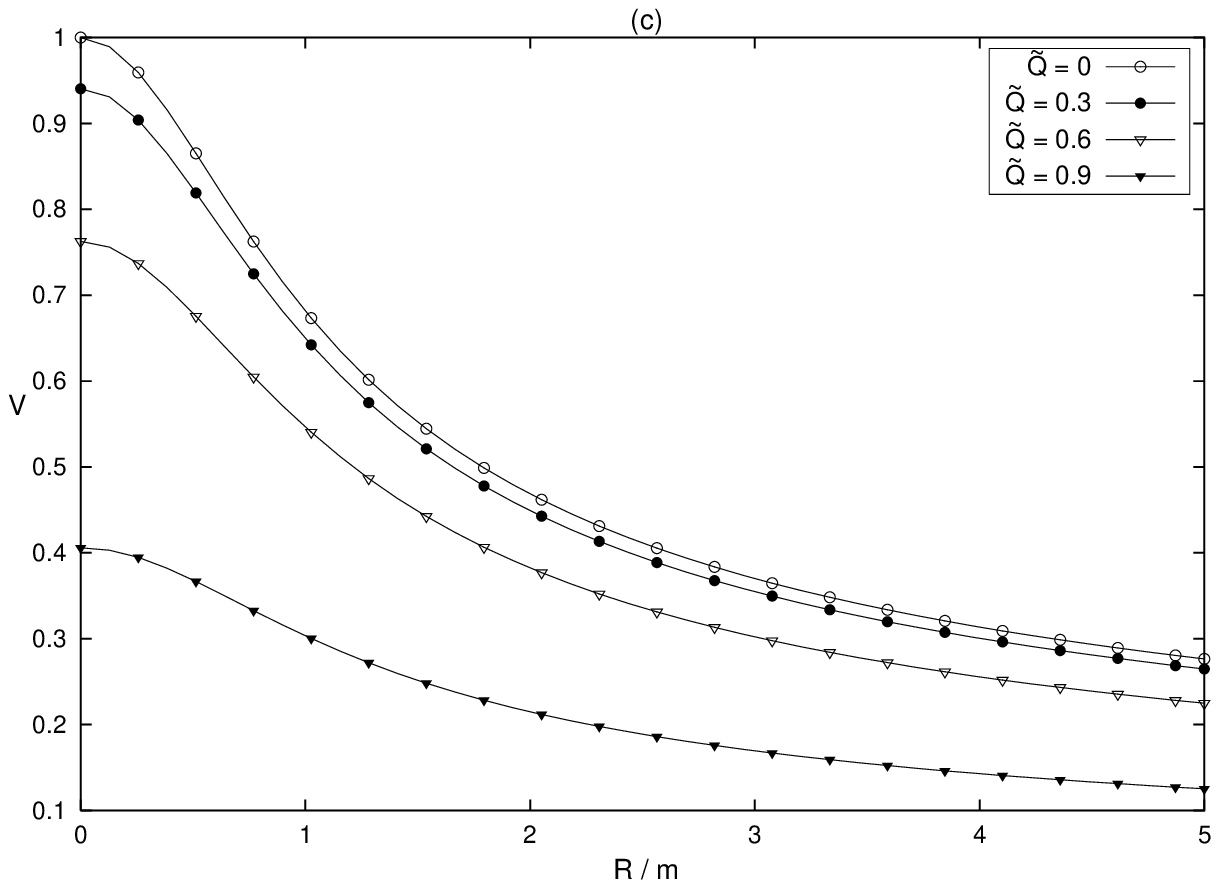}
\includegraphics[scale=0.6]{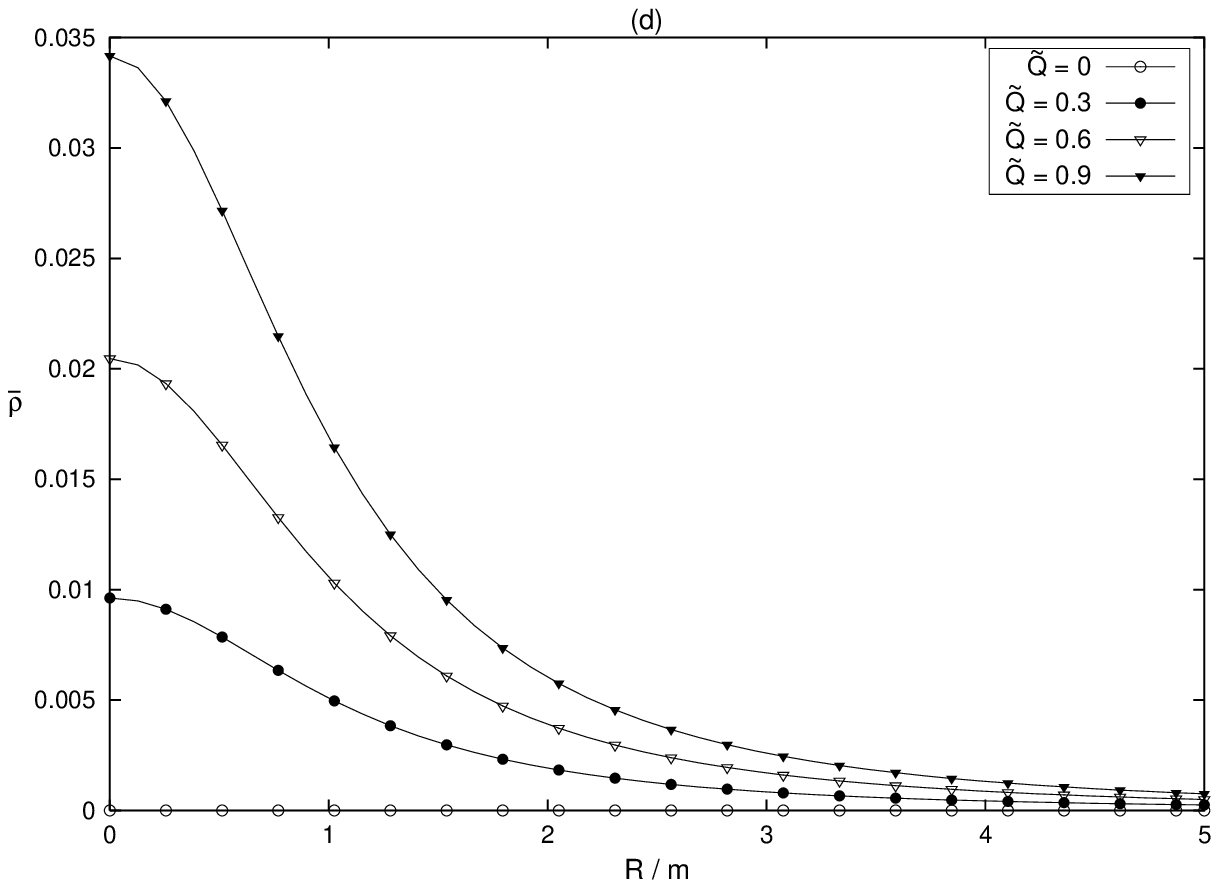}

\caption{(a) The surface energy density $\bar{\sigma}$, Eq.\ (\ref{eq_sigma_r}), (b) the
pressure $\bar{P}$, Eq.\ (\ref{eq_P_r}), (c) the sound velocity $V$ Eq.\ (\ref{eq_V_r}) and
(d) the charge density $\bar{\rho}$, Eq.\ (\ref{eq_rho_r}) for the disk with $\tilde{a}=1$,
$\tilde{Q}=0$, $0.3$, $0.6$ and $0.9$ as functions of $\tilde{R}$.} \label{fig2}
\end{figure}

Fig.\ \ref{fig2} (a)--(d) shows, respectively, the surface energy density $\bar{\sigma}$, pressures
$\bar{P}$, sound velocity $V$ and charge density $\bar{\rho}$ with $\tilde{a}=1$, $\tilde{Q}=0$, $0.3$,
$0.6$ and $0.9$ as functions of $\tilde{R}$. As charge increases, the disks become less 
relativistic for the same cut parameter; energy density and pressures are lowered and charge 
density becomes more concentrated near the disk center.

\begin{figure}
\centering
\includegraphics[scale=0.6]{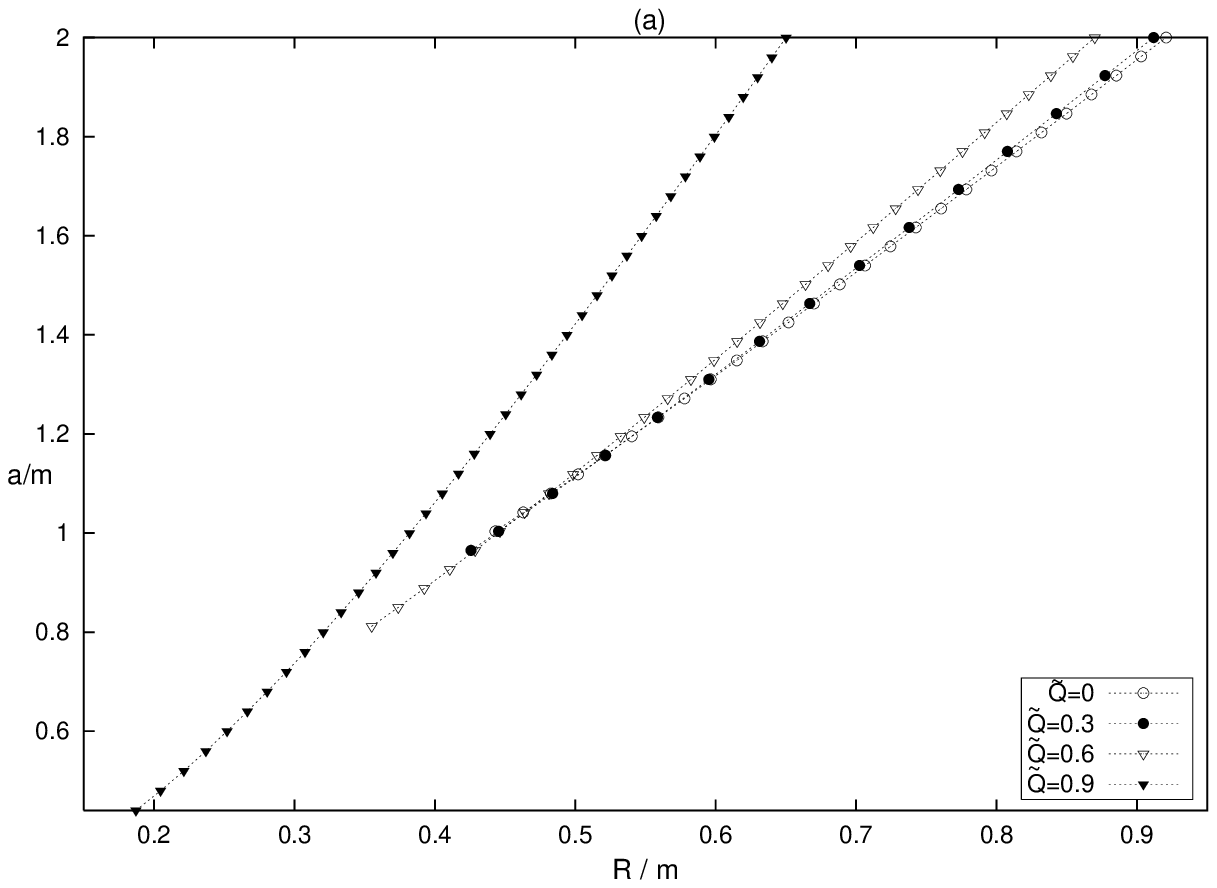}
\includegraphics[scale=0.6]{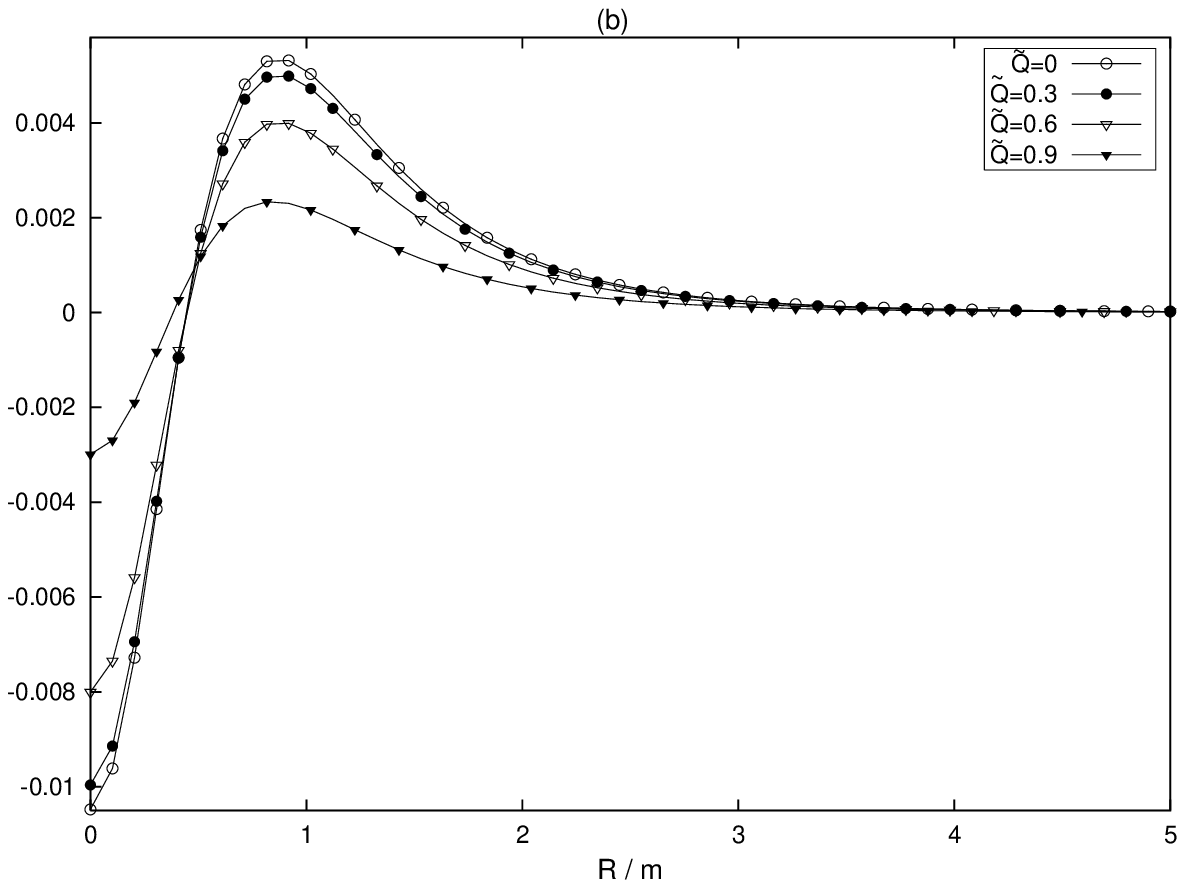}
\caption{(a) Curves where Eq.\ (\ref{eq_est_fluid}) defining
 stability changes sign as functions of radius $\tilde{R}$,
cut parameter $\tilde{a}$ and charge $\tilde{Q}$. Disks are stable at the left of each curve. (b) Curves of the left 
side of Eq.\ (\ref{eq_est_fluid}) for the same parameters as in Fig.\ \ref{fig2}.} \label{fig3}
\end{figure}

Fig.\ \ref{fig3}(a) is a graph of curves where Eq.\ \ref{eq_est_fluid} changes sign. The curves
have been plotted only for ranges of parameter $\tilde{a}$ where $V^2<1$ (Fig.\ \ref{fig1}). At the left of
each curve, stability condition (\ref{eq_est_fluid}) is satisfied. Thus the disks are stable only in a small
region near their centers.  We also note that the charge decreases the radii of stability. The left side of
Eq.\ \ref{eq_est_fluid} is plotted in Fig.\ \ref{fig3}(b) for the same parameters as in Fig.\ \ref{fig2}.

\section{Discussion} \label{sec_disc}

We applied the ``displace, cut and reflect'' method on the Reissner-Nordstr\"{o}m solution
in isotropic coordinates and generated static charged disks made of perfect fluid. We also 
derived a simple criteria for the stability for charged and uncharged perfect fluid disks that is an extension
of the Rayleigh criteria of stability for rotating fluids. The addition of charge decreases the energy density and pressures
near the disk's centers, while charge density in enhanced there. Furthermore, presence of charge 
decreases the stable regions of the disks, that are reduced to small regions near the center even in the absence of charge.
This is a rather different result from that of our previous stability analysis of the uncharged perfect fluid 
disk \cite{Vogt} based on the stability study of circular orbits of test particles, where we found that the disks were completely
stable for $\tilde{a}\gtrapprox 1.016$. It is worthwhile to note that the criteria used to study stability of 
disks made of counterrotating matter is based on a particle consideration, whereas the stability criteria 
derived for perfect fluid disks is based on collective phenomena. Therefore, the stability criteria derived 
in this paper seems to us to be more appropriate for the study of perfect fluid disks.

In principle it is possible to add charged halos to the disks presented in this work by applying the
``displace, cut and reflect'' method to a static charged sphere of perfect fluid in isotropic coordinates 
(see the Appendix). The halos could have charge of the same sign or opposite sign of the disk's charge
such that the entire object could be neutral or have an arbitrary charge. Although there exist many exact solutions for charged fluid
spheres in Schwarzschild coordinates (see, for example, \cite{Ivanov}), we have not found in the literature similar solutions
in isotropic coordinates. This may be due to the fact that in these coordinates the Einstein-Maxwell equations for a charged perfect fluid are reduced to
 a system of highly nonlinear
coupled second order differential equations for both metric functions
$\nu (r)$ and $\lambda (r)$ and for the electrostatic potential $\phi (r)$, so
the task of finding exact solutions is more involved than in canonical spherical coordinates. We believe that this search for exact solutions in isotropic coordinates is worthwile.

\begin{acknowledgments}
D.\ V.\ thanks CAPES for financial support. P.\ S.\ L.\ thanks FAPESP and CNPq for financial support.
\end{acknowledgments}

\appendix*
\section{}

\begin{figure}
\centering
\includegraphics[scale=0.7]{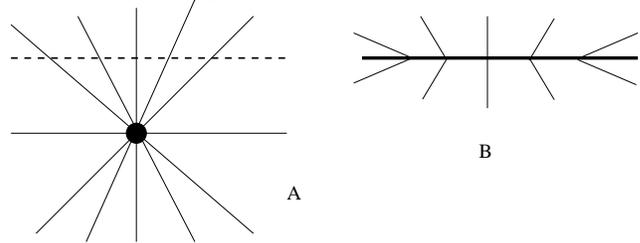}
\caption{An illustration of the ``displace, cut and reflect'' method for the generation of disks.
In A the spacetime with a singularity is displaced and cut by
 a plane (dotted line), in B the part with singularities is disregarded and the
upper part is reflected on the plane.} \label{fig_schem1}
\end{figure}

\begin{figure}
\centering
\includegraphics[scale=0.7]{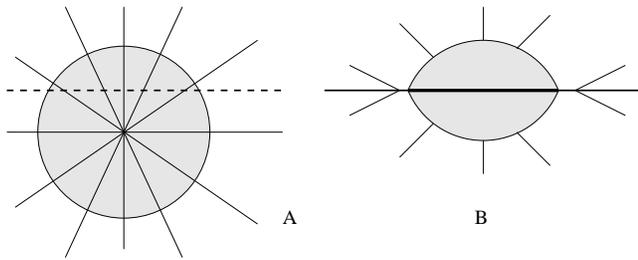}
\caption{An illustration of the ``displace, cut and reflect'' method for the generation of charged disks with halos. In A the sphere of perfect fluid is displaced and cut by
 a plane (dotted line), in B the part that contains the center of the sphere is disregarded and the
upper part is reflected on the plane.} \label{fig_schem2}
\end{figure}

In this appendix we give an overview of the ``displace, cut and reflect'' method used to generate the
metric and its material and electric content from a known solution of the Einstein-Maxwell field equations. 
The method can be divided
in the following steps that are illustrated in Fig.\ \ref{fig_schem1}:
First, in a space wherein we have a compact source of gravitational field,
 we choose a surface (in our case, the plane $z=0$) that divides the  space in
two pieces:  one with no singularities or sources and the other with the
 sources. Then we
disregard the part of the space with singularities and use the surface to make an inversion of the
nonsingular part of the space. This results in a space with a
singularity that is a delta function with support on $z=0$.
The same procedure can be used with a static sphere of charged perfect fluid to generate charged 
disks with halos, as depicted in Fig.\ \ref{fig_schem2}: the sphere is displaced and cut by a distance 
$a$ less then its radius. The part of the space that contains the center of the sphere is disregarded. After 
the inversion of the remaining space, one ends up with a charged disk surrounded by a cap of 
charged perfect fluid.
If the internal fluid solution is matched to the Reissner-Nordstr\"{o}m metric Eq.\ (\ref{eq_reiss2}), the
outer part of the disk will have the physical properties deduced in Sec.\ \ref{sec_disk}, while the 
properties of the inner part will depend on the particular fluid solution.

\end{document}